# Intrinsic ionic screening of the ferroelectric polarization of KTP

Mario Maglione[1], Anand Theerthan[1], Vincent Rodriguez[2],

Alexandra Peña[3,4], Carlota Canalias[5], Bertrand Ménaert[3,4], Benoît Boulanger[3,4]

…………………

[1] ICMCB-CNRS, UPR 9048, Université de Bordeaux, 87 av. Schweitzer, Pessac Cedex 33608, France

[2] Institut des Sciences Moléculaires, CNRS UMR 5255, Université de Bordeaux, 351 cours de la Libération, Talence Cedex 33405, France

[3] Univ. Grenoble Alpes, Inst NEEL, F-38042 Grenoble, France

[4] CNRS, Inst NEEL, F-38042 Grenoble, France

[5] Departement of Applied Physics, Royal Institute of Technology, Roslagstullsbacken 21, 10691 Stockholm, Sweden

**Abstract**

Mobile charges and lattice polarization interact in ferroelectric materials because of the Coulomb interaction between the mobile free charges and the fixed lattice dipoles. We have investigated this mutual screening in KTiOPO$_4$, a ferroelectric/superionic single crystal in which the mobile charges are K$^+$ ions. The ionic accumulation close to the crystal surfaces leads to orders of magnitude increase of the Second Harmonic Generation. This ionic space charge model is supported by the absence of such an effect in non-ionic conductor but ferroelectric BaTiO$_3$, by its temperature dependence in KTiOPO$_4$ and by its broad depletion at domain walls.

Keywords:

Corresponding author: Mario Maglione maglione@icmcb-bordeaux.cnrs.fr

The interplay between long range ionic mobility and polarization in ferroelectrics is increasingly investigated. It is indeed very interesting to understand the basic physics behind the coexistence of such two mutually exclusive properties [1] and maybe to achieve breakthrough functionalities, as: large conductivity [2], photovoltaics [3,4], electro-catalysis[5], electrochromism [6]. Most of the experiments up to date rely on thin films and crystal surfaces using near field set ups for which the tip-enhanced electric field is the main driving force for ionic migration [7]. In the present report, we want to investigate the spontaneous ionic accumulation at ferroelectric crystal surfaces at larger scale with no need for a large electric field. To this aim, we choose a Non Linear Optical (NLO) single crystal which both displays ferroelectricity and large superionic conductivity.

NLO single crystals are used as passive components in solid state lasers. The second order electric susceptibility coefficients enable Second Harmonic Generation (SHG) resulting in optical frequency conversion, for example from Infra-Red to visible range. Aside from their net polarization which is mandatory for SHG, some of the most efficient NLO crystals are good ionic conductors. Let us mention a partial list including Lithium Niobate ($LiNbO_3$), Potassium Dihydrogen Phosphate ($KH_2PO_4$, KDP) and Potassium Titanyl Phosphate ($KTiOPO_4$, KTP). In the former case, $Li^+$ is the key ion contributing in both the polarization and the ionic conductivity, while it is $H^+$ in the second compound and $K^+$ in the last one. In all these cases, the interplay between ionic conductivity and polarization is used to fix the polarization state using controlled ionic diffusion [8]. In the search for further evidence of actual coupling between polarization and ionic conductivity, we have recently shown that the piezoelectric properties are well altered at the right temperature where ionic conductivity ceases [9]. We linked the observed splitting of piezoelectric resonance peaks to the accumulation of $K^+$ on each side of the crystals thanks to the Coulomb interaction between the slowing down ions and the lattice polarization in KTP. This was however a very schematic view that required further experimental support. In the present report, we look for an actual mapping of such space charges in KTP that are detectable through the electro-optic (EO) effect using second harmonic (SHG) microscopy.[10,11] For that purpose, we have used a SHG microscope operating in the reflection mode, thus enabling a straightforward local mapping in the 3 spatial directions with a microscale resolution since the effect of the phase mismatch for a backward SH wave cannot be effective in that case. We find a sharp decrease of the effective net SHG activity from the near surface towards the inner crystals over a few tens of microns. The inner base coefficients are compatible with the well-known SHG parameters of KTP. We thus ascribe the near surface enhancement of the net SHG signal to an EO effect produced by the accumulation of $K^+$ ions. We have checked and confirmed that no such additional contribution occurs non-ionic conductor but ferroelectric $BaTiO_3$. These findings are a direct confirmation of the intrinsic ionic screening of ferroelectric polarization while it usually occurs through external charges, as electrons or charged radicals for example, in all investigated ferroelectrics.

SHG microscopy is an efficient way for probing local polarization in many instances when the macroscopic polarization does not perfectly reflects the local polar state. This was clearly evidenced in the case of ferroelectrics, relaxors and dipolar glasses [12, 13] where the local chemistry and structure differs from the average one. It was also applied with a great success to the mapping of ferroelectric / ferromagnetic domains in numerous multiferroic materials [14, 15]. High resolution SHG microscopy was also used for investigating the spatial switching of the ferroelectric polarization at domain walls [8, 16]. These previous investigations were achieved using a SHG microscope in the transmission mode providing two-dimensional maps while integrating over the propagation direction. Three-dimensional mapping of polarization through non-destructive SHG is thus to be applied in ferroelectric materials which are at the same time ionic conductors.

The typical micro-SHG experiment is basically very simple. The incident light with frequency ω is focused on the sample using a microscope objective and scanned across the sample. The sample can be imaged with a diffraction-limited lateral resolution of the order of 300-500 nm for a fundamental wave in the near-IR range ~ 1000 nm. The SHG signal (at double frequency 2ω) can be collected in the transmission mode where coherent (phase) processes occur efficiently giving thus an intense forward SHG signal [11]. By contrast, the signal in the reflection mode is likely incoherent and thus definitively weaker since phase relations are confined to typical spatial dimensions much smaller than the size of the laser spot. A main advantage of SHG microscopy is its unique capability to provide 3D images of the materials with micrometer resolution.

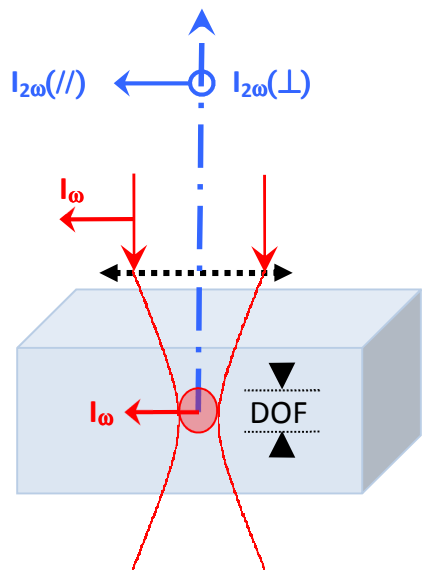

**Figure 1:** Orientation conventions for the backscattered parallel (//) and perpendicular (⊥) polarized light intensity $I_{2\omega}$ with respect to the incident light intensity $I_\omega$. Note here that the depth of focus (DOF) corresponds to the axial resolution estimated to 2-3 µm using an 100X objective with NA=0.50

The micro-SHG setup we used for this study is based on a modified micro-Raman spectrometer (Horiba HR800) allowing the analysis of backscattered light. This original setup gives the opportunity to perform µ-SHG [17-20] and also µ-Raman to get a direct link between physical properties and local structure [21-23]. The source is a diode-pumped picosecond laser (EKSPLA PL2200: pulse duration 65 ps, repetition rate 2kHz) operating at $\lambda$=1064 nm. The polarized incident laser beam is focused onto the sample using a 100x Mitutoyo Plan APO near-IR objective (infinity corrected with NA = 0.50). The energy per pulse is adjusted by a power unit composed of a rotating half-wave plate in front of a GLAN Taylor prism and it can be tuned (from 10 µJ to less than 10 nJ) notably at low energy to avoid eventual sample photo-degradation. The measurement of the SHG intensity $I_{2\omega}$ can be performed using two different light polarization configurations, namely the parallel (//) and perpendicular (⊥) configurations, corresponding to incident and scattered lights with collinear or orthogonal polarizations respectively (Figure 1). Since the polarization of KTP is predominantly along z direction, we used the // polarization configuration. The lateral spatial resolution is estimated to be 500 nm from the far field resolution limit. The depth of focus (DOF) was estimated to be ca. 2-3 μm. If the probed thickness is defined by the DOF the SHG intensity (I2ω) is related to the NLO susceptibility $d_{eff}$ as follows.

$$(I_{2\omega}) \propto DOF^2 * (d_{eff})^2 * (I_\omega)^2$$

Hence, keeping the same experimental conditions, the collected backward SHG signal is simply proportional to the square of the effective SHG susceptibility *$d_{eff}$*. In that case which is similar to Kurimura et al. [12], the collected backward SHG signal at high NA will not give any phase information but only a 3D map of the magnitude (loss of the sign) of the effective SHG response.

We first consider $BaTiO_3$ crystals as a reference non-ionic conducting ferroelectric material. As shown in figure 2a, surface domain walls are readily observed by standard optical microscopy. Keeping the crystal on the same stage but shifting to SHG microscopy and zooming at the black square shown in figure 2a, the same domains of about 20μm width, but with alternated polarization, display an SHG intensity *ca* 2000 in arbitrary unit (fig 2b). Clearly, on the crystal surface we observe with polarized SHG 180° and 90° walls as reported in linear optical investigations [24] and sketched on figure 2c. The only SHG intensity is recorded when the laser polarization matches the domain one. The domain wall show no SHG in the detection limit of our experiment. This means that the polarization at domain walls is negligible and that the 180° polarization switching takes place over distances much smaller than our spatial lateral resolution of ~0.5 μm. Using surface SHG as a reference, we further performed in depth SHG investigations which showed at some 60 μm depth [25] much larger SHG intensity *ca* 15000, *i.e.* more than 7 times the surface response. This is ascribed to deep domains having full matching with the input laser polarization as schematically sketched in figure 2e. All domains above the matched ones display marginal SHG for two possible reasons: at first only half of the volume of the DOF is concerned when focusing at the interface, and secondly the surface signals of figure 2b is probably a mixture of the domains that extend in-depth as reported in Fig. 2e. Again for the domains of highest efficiency, the domain wall SHG is negligible compared with the bulk domain one. Such experiments were repeated at several locations on $BaTiO_3$ single crystals: we confirmed that no specific surface SHG appeared, and the main SHG activity is strictly resulting from the polarization alignment.

(a)

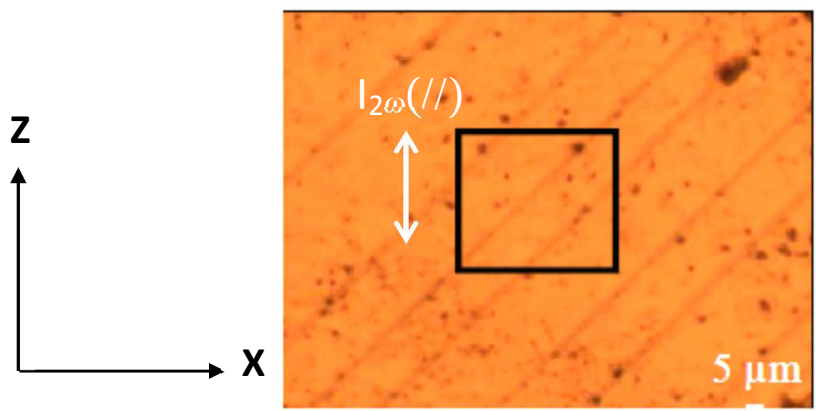

(b)                                                                 (c)

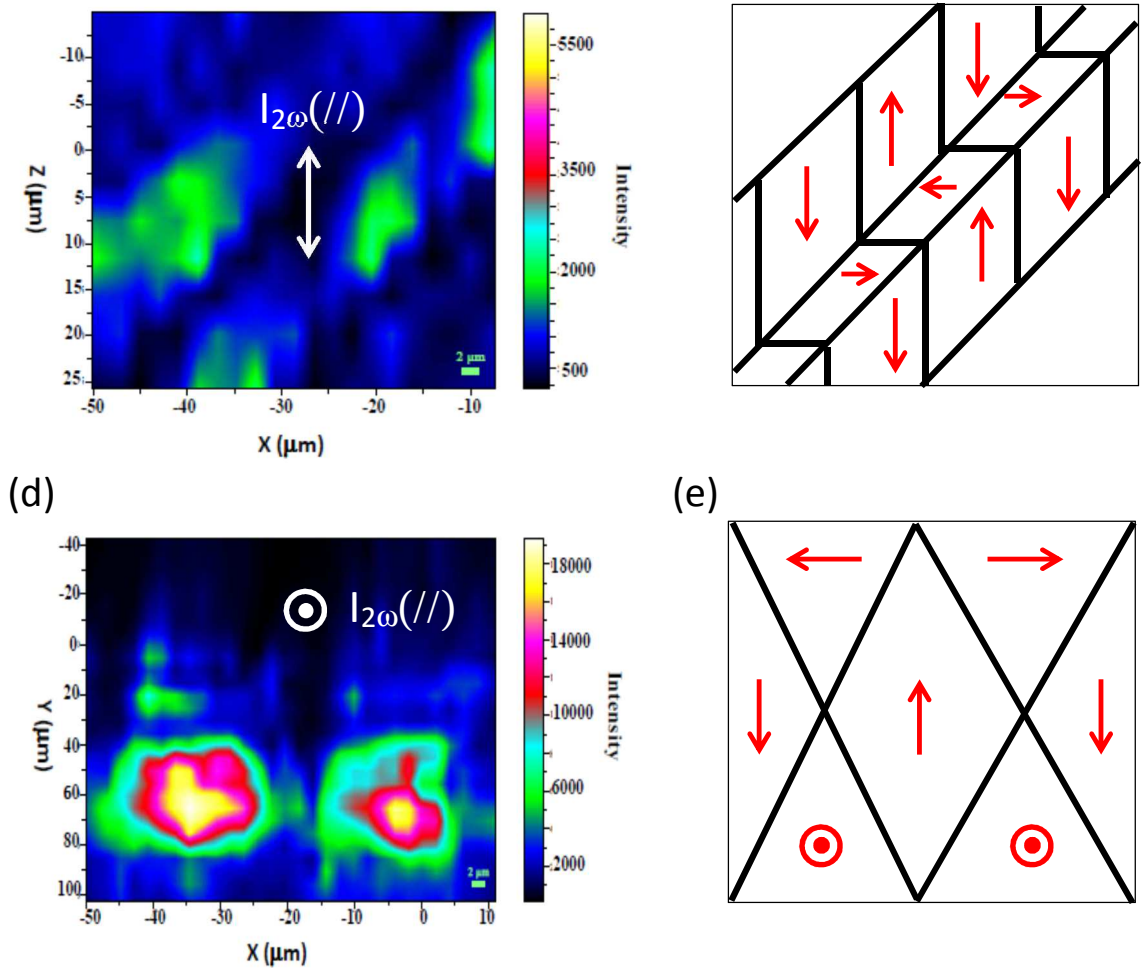

**Figure 2:** BaTiO3 single crystal surface seen by optical microscopy (a) and by SHG microscope (b) evidencing alternated vertical polarized domains of about 20µm width each (c). The in-depth maps (d) shows that the surface SHG signal (max 2000) is much increased to 15000 at depth of 60µm where the polarization matches the domain orientations with possible corresponding polarization (e).

This BaTiO3 polarized µ-SHG investigation will be taken as a reference for the further investigation of KTP single crystals. These two crystals being uniaxial ferroelectrics, we set the polarization axis of the crystal as parallel to the laser polarization for the whole experiment to be performed under polarization matching conditions for any position inside the crystal. On figure 3b, near surface SHG intensity is observed but without any domain features, which is in agreement with the experimental geometry (figure 3a). This map was recorded 10 micrometers beneath the surface [25] in order to remove the contributions from surface defects. As evidenced from the grey scale (color online), the SHG intensity lies in between 500 and 800 close to room temperature (figure 3a). We offer no explanation for the in plane intensity fluctuations which should not be seen in case of homogeneous ferroelectric polarization. Because of the uniaxial polarization of KTP and because SHG is sensitive only to the direction and not to the polarization orientation, ferroelectric domain should contribute equally. On cooling below 200K, a significant decrease of the overall intensity towards less than 300 occurs (figure 3c-3e). We ascribe such decrease to the super-ionic transition that takes place in KTP at 200K [26]. This temperature evolution is an evidence that mobile K$^+$ ions do contribute to the SHG signal, which could in turn explain the in-plane SHG fluctuations.

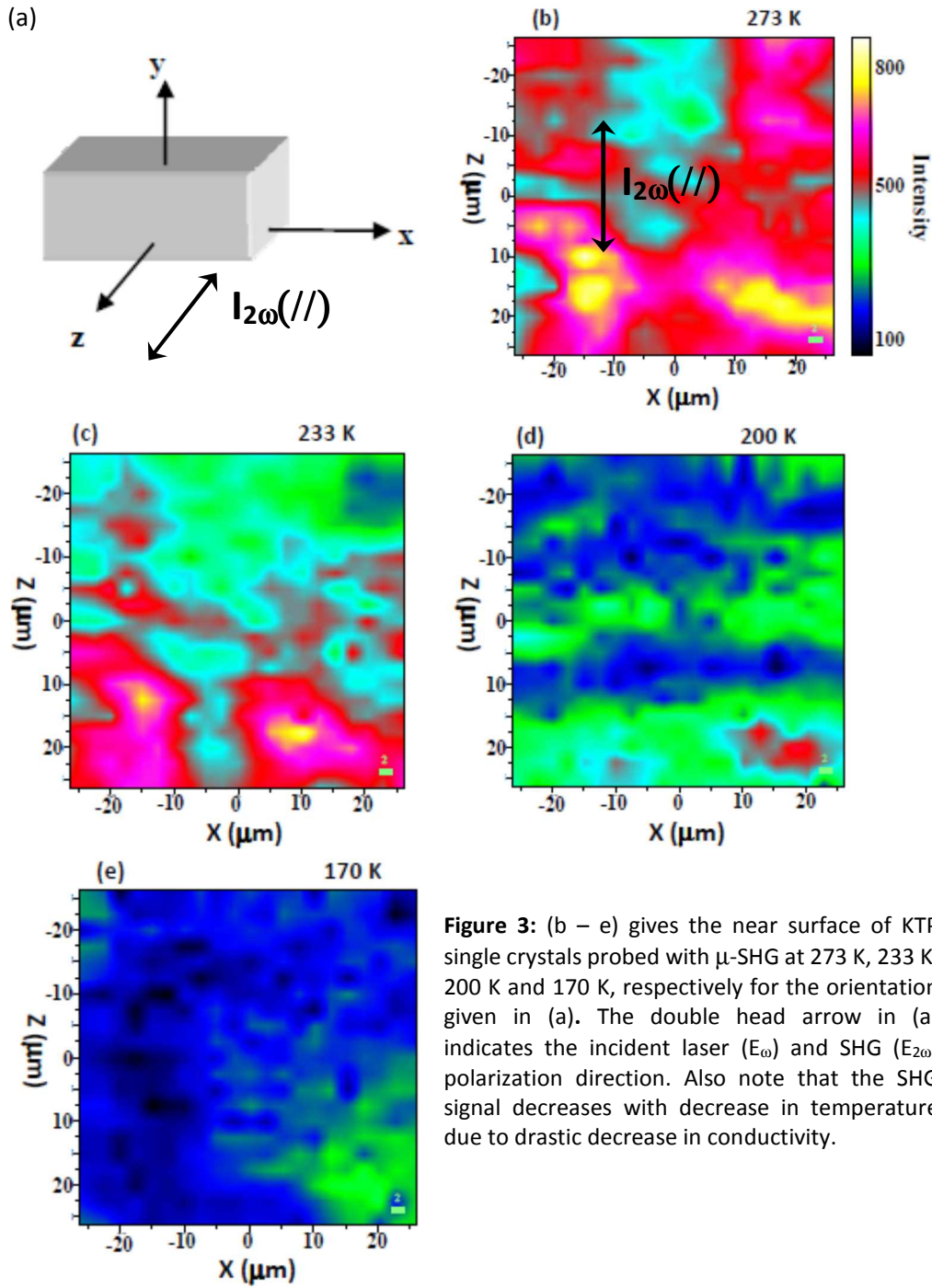

**Figure 3:** (b – e) gives the near surface of KTP single crystals probed with µ-SHG at 273 K, 233 K, 200 K and 170 K, respectively for the orientation given in (a). The double head arrow in (a) indicates the incident laser ($E_\omega$) and SHG ($E_{2\omega}$) polarization direction. Also note that the SHG signal decreases with decrease in temperature due to drastic decrease in conductivity.

We also observed well defined evolutions of the piezoelectric resonances in KTP at 200K. Since SHG and piezoelectric parameters are both linked to the same polarization state, SHG thus confirms that KTP overall polarization state changes drastically at 200K. We suggested a very schematic model for the piezoelectric parameters changes at 200K, by assuming that mobile $K^+$ ions and lattice polarization are strongly connected and thus surface accumulation of $K^+$ may occur thanks to the Coulomb interaction [9]. We performed in depth SHG analyses to get insight into this assumed ionic accumulation and its possible contribution to the ferroelectric polarization. On figure 4a, the SHG intensity (between 400 and 600) is localized at 20 to 30 µm beneath the surface. We note that the base in-depth intensity that appears dark on this map gives the same SHG activity as the one reported in literature [27, 28]. This means that the near surface SHG is a very large amplification of these bulk coefficients by more than 2 orders of magnitude.

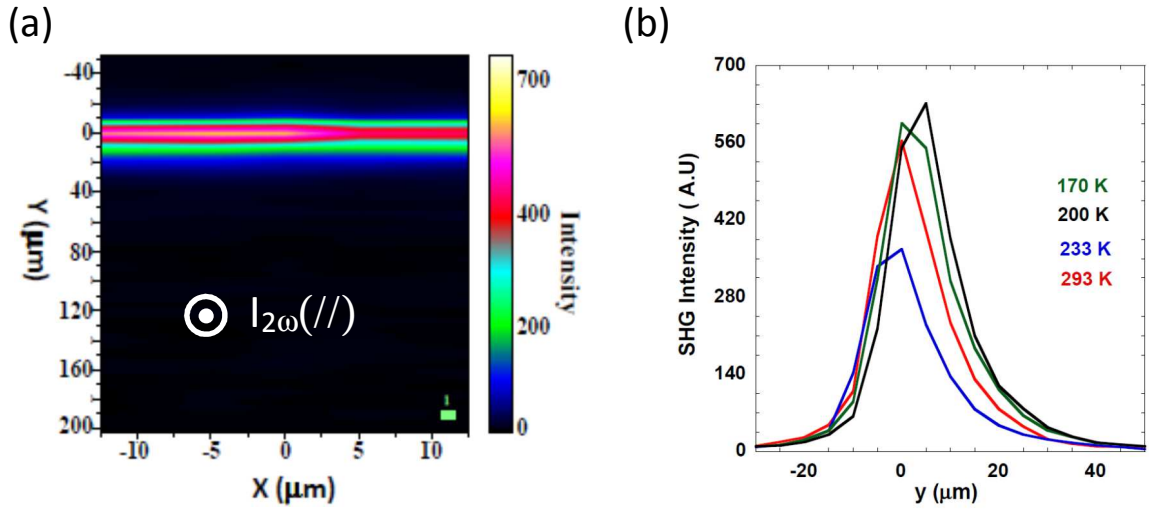

**Figure 4**: (a) in depth mapping of SHG intensity in KTP at 293K and (b) cross sectional intensity at a fixed x versus the depth y. Some thermal variation of these peaks occurs with a sharp minimum close to 230K, in the temperature range where ionic mobility decreases.

We recorded such in-depth maps *versus* temperature from 298K down to 170K. Figure 4b shows the profile of average SHG intensity along *y* direction (depth) obtained from the *xy* cross-sectional images. The SHG intensity at the surface, *i.e.* $y = 0$ µm, is high at any temperature, the width of the peak giving the thickness of the space charge layer. At room temperature, the SHG intensity is high, and at 233 K it reaches a minimal amplitude. This minimum coincides exactly with the decrease in conductivity that starts to happen at this temperature range. At 200 K and 170 K, the SHG intensity increases again, which can be explained on the basis of the intrinsic polarization of KTP becoming more pronounced at low temperatures as seen from pyroelectric measurements [9]. Here it is worth noting that at those optical frequencies of the fundamental and harmonic waves, *i.e.* 1064 nm and 532 nm, SHG is sensible to electronic fluctuations (in the range of fs to ps) mostly due to the Ti-O bonds (fs range) in KTP. These electronic fluctuations relate not only the instantaneous electronic polarization but also fluctuations of nucleus motions (ps range) that induce also electronic fluctuations as it happens in Raman scattering for example. Then, the overall high SHG intensity at the surface is a consequence of the static inhomogeneous distribution of potassium ions (static space charge) but also of their dynamical hopping (dynamical contribution not solved at optical frequencies), which gives rise to electron density fluctuation. This kind of static space charge layer due to migration of ions has been observed even in glasses.

To confirm that the near surface SHG intensity is indeed related to the polarization state of the crystals, we performed the same in depth mapping of Periodically Poled KTP (PPKTP). In such crystals, alternated domains of reverse relative orientation are generated with a periodicity of about 20µm to achieve a quasi-phase matching for improved and tunable frequency conversion. We show on figure 5 the in plane and the in depth map of PPKTP.

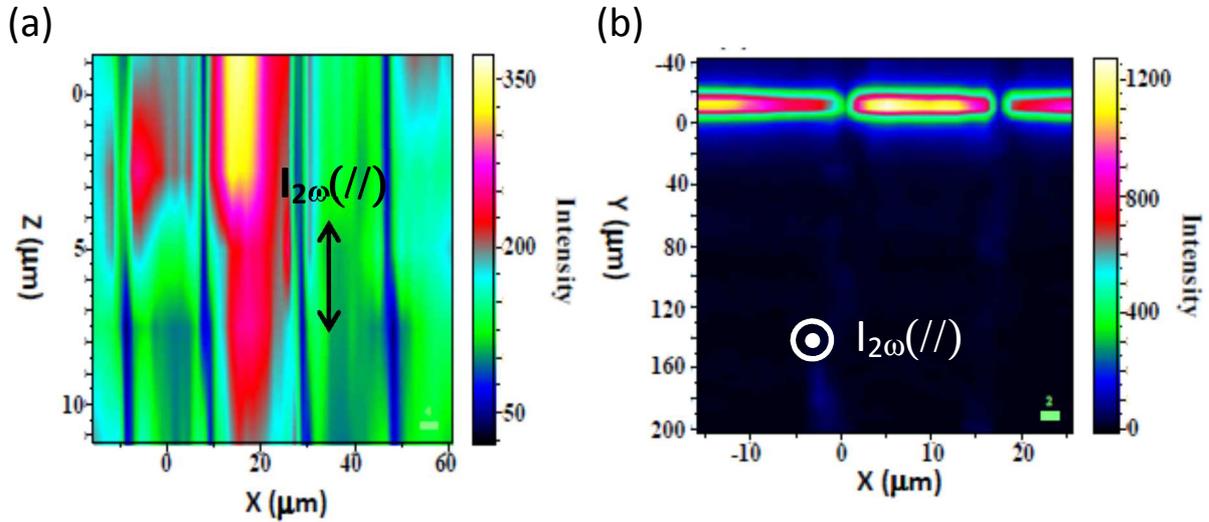

**Figure 5:** polarized SHG maps of periodically poled KTP crystal at 298K. The surface map (a) clearly evidences the 20µm-periodicity of the alternated domains with a sharp decrease at all domain walls. The in-depth map (b) confirms this periodicity and the large enhancement of SHG at the near surface for each domain. The background intensity in the inner crystal is compatible to the usually reported SHG of KTP.

The surface map clearly evidences the alternatively oriented domains of individual width 20µm, which is in perfect agreement with the targeted periodicity of these Periodically Poled single crystals. We offer no explanation for the intensity fluctuations of this surface map, but spurious species adsorbed to this un-cleaned surface maybe anticipated. The domain walls display no SHG polarization as expected from the absence of polarization between two domains. On the in-depth mapping (fig 5b), the strong enhancement of SHG within a depth of 10µm similar to the unpoled crystals is observed. As seen from the intensity scale, the maximum SHG intensity within the domains if 4 times higher than the surface SHG (fig 5a) and orders of magnitude larger than the in-depth SHG. As for unpoled crystals, this background SHG is the reported one for KTP single crystals. There is thus a very strong SHG amplification at the near surface. The absence of above-background SHG at the domain walls shows that surface defects are unlikely to contribute to the inner domains signal. It thus arises from the polarization state of KTP that is compensated at the domain walls. We again ascribe this amplification to the ionic charge accumulation resulting from the screening of the ferroelectric polarization. Since such polarization is alternatively pointing outward and inward, then such ionic space charge is alternatively an accumulation and a depletion of $K^+$ ions. The effective ionic polarization is thus alternatively positive and negative, but such sign oscillations have no effect on the SHG activity since it is sensitive to the square of the polarization. An ionic space charge model can also account for the thickness of the SHG depletion at the domain wall: it is of the order of 1µm, and orders of magnitude larger than the structural thickness of 1 atomic plane [26]. At any interface the free charges accumulation/depletion is much larger than the interface morphology itself, and we assume that it is what happens at domain walls in KTP.

We thus have provided evidence for a strong SHG amplification near the surface of KTP that is simultaneously a ferroelectric and an ionic conductor. This was found in single domain as well as in periodically poled crystals with no signal at the domain boundaries. These observations rule out extrinsic or defects related origin for the surface SHG. No such amplification could be found in $BaTiO_3$ where only the matching between the laser and lattice polarizations leads to measurable SHG. We thus concluded that ionic motion and accumulation as space charges close to the surface can provide the necessary extra-polarization for SHG amplification in KTP. In-depth chemical analysis is underway to confirm such an ionic accumulation in KTP

## Acknowledgments

V. R. thanks F. Adamietz for technical assistance and the Région Aquitaine for financial support in optical, laser, and computer equipment.